\RequirePackage{ifpdf}
\documentclass[12pt,letterpaper]{JHEP3}
\pdfoutput=1
\usepackage{epsfig}
\usepackage{subcaption}
\usepackage{graphicx}

\usepackage[english]{babel}
\usepackage{mathtools}

\usepackage{cite}

\def\be{\begin{equation}}
\def\ee{\end{equation}}

\def\be{\begin{equation}}
\def\ee{\end{equation}}

\DeclarePairedDelimiter\floor{\lfloor}{\rfloor}
\allowdisplaybreaks

\graphicspath{{Figures/}}

\title{Conditionally Extended Validity of Perturbation Theory: Persistence of AdS Stability Islands}

\author{Fotios Dimitrakopoulos and I-Sheng Yang\\
IOP and GRAPPA, Universiteit van Amsterdam, \\
Science Park 904, 1090 GL Amsterdam, Netherlands
}

\abstract{Approximating nonlinear dynamics with a truncated perturbative expansion may be accurate for a while, but it in general breaks down at a long time scale that is one over the small expansion parameter. There are interesting occasions in which such breakdown does not happen. We provide a mathematically general and precise definition of those occasions, in which we prove that the validity of truncated theory trivially extends to the long time scale. This enables us to utilize numerical results, which are only obtainable within finite times, to legitimately predict the dynamic when the expansion parameter goes to zero, thus the long time scale goes to infinity. 

In particular, this shows that existing non-collapsing solutions in the AdS (in)stability problem persist to the zero-amplitude limit, opposing the conjecture by Dias, Horowitz, Marolf and Santos that predicts a shrinkage to measure-zero\cite{DiaHor12}. We also point out why the persistence of collapsing solutions is harder to prove, and how the recent interesting progress by Bizon, Maliborski and Rostoworowski is not there yet\cite{BizMal15}.
}

\begin{document}

\section{Introduction and Summary}

\subsection{Truncated Perturbative Expansion}

A linear equation of motion $D\phi=0$ often has close-form analytical solutions. A nonlinear equation, $D\phi = F_{\rm nonlinear}(\phi)$, on the other hand, usually does not. One can attempt to expand $F_{\rm nonlinear}$ when $\phi$ is small. For example,
\begin{equation}
D\phi = F_{\rm nonlinear}(\phi) = \phi^2 + \mathcal{O}(\phi^3)~.
\label{eq-truncate}
\end{equation}
When the amplitude is small, $|\phi|<\epsilon\ll 1$, one can solve the truncated equation of motion that includes the $\phi^2$ term as a perturbative expansion of $|\phi^2/\phi| < \epsilon$ from the linear solutions. For a small enough choice of $\epsilon$, this can be a good enough approximation to the fully nonlinear theory. Unfortunately, this will only work for a short amount of time. After some time $T\sim\epsilon^{-1}$, the correction from the first nonlinear order accumulates and becomes comparable to the original amplitude. Thus the actual amplitude can exceed $\epsilon$ significantly to invalidate the expansion.\footnote{The notion of ``time'' here is just to make connections to practical problems for physicists. The general idea is valid whenever one tries to solve perturbation theory from some limited boundary conditions to a far-extended domain.}

A slightly more subtle question arises while applying such a truncated perturbation theory. Occasionally, there can be accidental cancellations while solving it. Thus during the process, the amplitude may stay below $\epsilon$ for $T\sim\epsilon^{-1}$. {\it In these occasions, are we then allowed to trust these solutions?}

It is very tempting to directly answer ``no'' to the above question. When $T\sim\epsilon^{-1}$, not only the accumulated contribution from $\phi^2$, which the theory does take into account, modifies $\phi$ significantly.  The $\phi^3$ term that the theory discarded also modifies $\phi^2$, and so on so forth. Since we have truncated all those even higher order terms which may have significant effects, the validity of the expansion process seems to unsalvageably break down.

The above logic sounds reasonable but not entirely correct. In this paper, we will demonstrate that at exactly $T\sim\epsilon^{-1}$ time scale, the opposite is true. {\it These ``nice'' solutions we occasionally find in the truncated theory, indeed faithfully represent similar solutions in the full nonlinear theory.} This idea is not entirely new. We are certainly inspired by the application of the two-time formalism and the renormalization flow technique in the AdS-(in)stability problem, and both of them operate on this same concept \cite{BalBuc14,CraEvn14}.\footnote{We thank Luis Lehner for pointing out that some Post-Newtonian expansions to General Relativity also shows validity at this long time scale\cite{Wil14}.} However, one may get the impression from those examples that additional techniques are required to maintain the approximation over the long time scale. One main point of this paper is to establish that the validity of truncated theory extends {\it trivially} in those occasions. As long as the truncated theory is implemented recursively, which is the natural way to solve any time evolution anyway, it remains trustworthy in those occasions.\footnote{It is extremely likely that a capable mathematician can directly point to a textbook material to backup this claim. However, such reference is hard to find for us, and may not be very transparent to physicists. Furthermore, the actual mathematical proof is quite simple, so we will simply construct and present them in this paper. Any suggestion to include a mathematical reference is welcomed.}

In Section \ref{sec-thm1}, we will state and prove a theorem that guarantees a truncated perturbative expansion, implemented recursively, to approximate the full nonlinear theory accurately in the long time scale in the relevant occasions. More concretely, this theorem leads to the two following facts:
\begin{itemize}
\item If one solves the truncated theory and finds solutions in which the amplitude remains small during the long time scale, then {\it similar} solutions exist in the full nonlinear theory.\footnote{Be careful that sometimes, especially in gauge theories, the full nonlinear theory might impose a stronger constraint on acceptable initial conditions. One should start from those acceptable initial conditions in order to apply our theorem. We thank Ben Freivogel for pointing this out.}
\item If numerical evolution of the full nonlinear theory provides solutions in which the amplitude remains small during the long time scale, then a truncated theory can reproduce {\it similar} solutions.
\end{itemize}

Formally, the meaning of ``{\it similar}'' in the above statements means that the difference between two solutions goes to zero faster than their amplitudes in the zero amplitude limit.
This theorem provides a two-way bridge between numerical and analytical works. Anything of this nature can be quite useful. For example, numerical results are usually limited to finite amplitudes and times, while the actual physical questions might involve taking the limit of zero amplitude and infinite time. With this theorem, we can start from known numerical results and extend them to the limiting case with analytical techniques.

In Section \ref{sec-thm2}, we will prove another theorem which enables us to do just that in the AdS (in)stability problem. The key is that the truncated theory does not need to be exactly solved to be useful. Since its form is simpler than the full nonlinear theory, it can manifest useful properties, such as symmetries, to facilitate further analysis. Since it is a truncated theory, the symmetry might be an approximation itself, and na\"ively expected to break down at the long time scale. Not surprisingly, using a similar process, we can again prove that such symmetry remains trustworthy in the relevant occasions.

It is interesting to note that the conventional wisdom, which suggested an unsalvageable breakdown at $T\sim\epsilon^{-1}$, is not entirely without merits. We can prove that both theorems hold for $T=\alpha\epsilon^{-1}$ for an arbitrarily but $\epsilon$-independent value of $\alpha$. However, pushing it further to a slightly longer, $\epsilon$-dependent time scale, for example $T\sim\epsilon^{-1.1}$, the proofs immediately go out of the window. The situation for $T\sim\epsilon^{-1}(-\ln \epsilon)$ is also delicate and will not always hold. Naturally, for time scales {\bf longer} than $T\sim\epsilon^{-1}$, one needs to truncate the theory at an even higher order to maintain its validity. A truncated theory up to the $\phi^m$ term will only be valid up to $T\sim\epsilon^{1-m}$.

\subsection{The AdS (In)Stability Problem}

In Section \ref{sec-island}, we will apply both theorems to the AdS-(in)stability problem\cite{BizRos11,DiaHor11,deOPan12,Lie12,DiaHor12,Mal12,BucLeh12,BucLie13,Biz13,MalRos13,MalRos13a,BalBuc14,MalRos14,HorSan14,CraEvn14,BasKri14,DFLY14,Yan15,BasKri15}. Currently, the main focus of this problem is indeed the consequence of the nonlinear dynamics of gravitational self-interaction, at the time scale that the leading order expansion should generically break down. Some have tried to connect such break down to the formation of black holes and further advocate that such instability of AdS space is generic. In particular, Dias, Horowitz, Marolf and Santos made the {\bf stability island conjecture}\cite{DiaHor12}. {\it Although at finite amplitudes, there are numerical evidence and analytical arguments to support measure-nonzero sets of non-collapsing solutions, they claimed that the sets of these solutions shrinks to measure zero at the zero amplitude limit.}

Since the relevant time scale goes to infinity at the zero amplitude limit, such conjecture cannot be directly tested by numerical efforts. Nevertheless, by the two theorems we prove in this paper, it becomes straightforward to show that such conjecture is in conflict with existing evidence. The physical intuition of our argument was already outlined in \cite{DFLY14}, and here we establish the rigorous mathematical structure behind it. 
\begin{itemize}
\item Theorem I allows us to connect non-collapsing solutions\cite{DiaHor12,BucLie13,MalRos13,BalBuc14} to analogous solutions in a truncated theory, both at finite amplitudes.
\item Theorem II allows us to invoke the rescaling symmetry in the truncated theory and establish those solutions at arbitrarily smaller amplitudes. 
\item Using Theorem I again, we can establish those non-collapsing solutions in the full nonlinear theory at arbitrarily smaller amplitudes.
\end{itemize} 
{\it Thus, if non-collapsing solutions form a set of measure nonzero at finite amplitudes as current evidence implies, then they persist to be a set of measure nonzero when the amplitude approaches zero. Since the stability island conjecture states that stable solutions should shrink to sets of measure zero, it is in conflict with existing evidence.}

It is important to note that defeating the stability island conjecture is not the end of the AdS (in)stability problem. Another important question is whether collapsing solutions, which likely also form a set of measure nonzero at finite amplitudes, also persist down to the zero-amplitude limit. It is easy to see why that question is harder to answer. Truncated expansions of gravitational self-interaction, at least all those which have been applied to the problem, do break down at a certain point during black hole formation. Thus, Theorem I fails to apply, one cannot establish a solid link between the truncated dynamic to the fully nonlinear one, and the AdS (in)stability problem remains unanswered. 

In order to make an equally rigorous statement about collapsing solutions, one will first need to pose a weaker claim. Instead of arguing for the generality of black hole formation, one should consent with ``energy density exceeding certain bound'' or something similar. This type of claim is then more suited to be studied within the validity of Theorem I, and it is also a reasonable definition of AdS instability. If arbitrarily small initial energy always evolves to have finite energy density somewhere, it is a clear sign of a runaway behavior due to gravitational attraction.\footnote{It is then natural to believe that black hole formation follows, though it is still not guaranteed and hard to prove. For example, a Gauss-Bonnet theory can behave the same up to this point, but its mass-gap forbids black hole formation afterward\cite{BucGre14a}.}

Finally, we should note that the truncated theory is already nonlinear and may be difficult to solve directly. If one invokes another approximation while solving the truncated theory, such as time-averaging, then the process becomes vulnerable to an additional form of breakdown, such as the oscillating singularity seen in\cite{BizMal15}. Even if numerical observations in some cases demonstrate a coincident between such breakdown and black hole formation, the link between them is not yet as rigorous as the standard established in this paper for non-collapsing solutions.

\section{Theorem I: Conditionally Extended Validity}
\label{sec-thm1}

Consider a linear space $\mathcal{H}$ with a norm satisfying triangular inequality.
\begin{equation}
||x + y|| \leq ||x|| + ||y||~, \ \ \ 
{\rm for \ all} \ x,~y\in\mathcal{H}~.
\end{equation}
Then consider three smooth functions $L, f, g$ all from $\mathcal{H}$ to itself. We require that $L(x)=0$ if and only if $x=0$, and it is ``semi-length-preserving''.
\begin{equation}
||L(x)||\leq ||x||~.
\label{eq-SLP}
\end{equation}
Note that this condition on the length is at no cost of generality. Given any smooth function $\bar{L}$ meeting the first requirement, we can always rescale it to make it exactly length-preserving and maintaining its smoothness.
\begin{equation}
L(x)\equiv \frac{||x||}{||\bar{L}(x)||}\bar{L}(x)~, \ \ 
{\rm if} \ \ x\neq0~; \ \ \ 
L(x)\equiv0~, \ \ {\rm when} \ \ x=0~.
\end{equation} 

$f$ and $g$ are supposed to be two functions that within some radius $r<1$, they are both close to $L$ but even closer to each other.
\begin{enumerate}
\item Being close to $L$: $\forall ~ ||x||<r~,$
\begin{equation}
||f(x)-L(x)|| < a||x||^m, \ \ 
||g(x)-L(x)|| < a||x||^m, \ \ \ 
{\rm for \ some} \ \ a>0~, \ \ m>1~.
\label{eq-Lclose}
\end{equation}

And doing so smoothly: $\forall ~ ||x||,||y||<r$ and some $b>0$,
\begin{eqnarray}
& & \bigg|\bigg|[f(x)-L(x)]-[f(y)-L(y)]\bigg|\bigg| 
< b\cdot||x-y||{\rm Max}(||x||,||y||)^{m-1}~, 
\label{eq-Lsmooth} \\ \nonumber
& & \bigg|\bigg|[g(x)-L(x)]-[g(y)-L(y)]\bigg|\bigg| 
< b\cdot||x-y||{\rm Max}(||x||,||y||)^{m-1}~.
\end{eqnarray}
\item Even closer to each other: $\forall ~ ||x||<r~,$
\begin{equation}
||f(x)-g(x)|| < c||x||^l~, \ \ \ 
{\rm for \ some} \ \ c>0~, \ \ l>m~.
\label{eq-FGclose}
\end{equation}
\end{enumerate}

Since this is a physics paper, we shall make the analogy to the physical problem more transparent by an example. Choose a finite time $\Delta t$ to evolve the linear equation of motion $D\phi=0$, $L$ is given by $L[\phi(t)] = \phi(t+ \Delta t)$. Similarly, evolving the full nonlinear theory $D\phi = F_{\rm nonlinear}(\phi)$ leads to a different solution $\phi$ that defines $f[\phi(t)] = \phi(t+\Delta t )$, and $D\phi = \phi^2$ defines $g[\phi(t)] = \phi(t+\Delta t)$. Furthermore, the norm can often be defined as the square-root of conserved energy in the linear evolution, which satisfies both the triangular inequality and the semi-preserving requirement.

From this analogy, evolution to a longer time scale is naturally given by applying these functions recursively. We will thus define three sequences accordingly.
\begin{eqnarray}
f_0 = g_0 = L_0 = x~, \ \ \ L_n = L(L_{n-1})~, \ \ \ 
f_n = f(f_{n-1})~, \ \ \  g_n = g(g_{n-1})~.
\end{eqnarray}
We will prove a theorem which guarantees that after a time long enough for both $g_n$ and $f_n$ to deviate significantly from $L_n$, they can still stay close to each other. \\ \ \\ 

{\bf Theorem I:} For any finite $\delta>0$ and $\alpha>0$, there exists $0<\epsilon<r$ such that if $||f_n||<\epsilon$ for all $0\leq n <\alpha\epsilon^{1-m}$, then $||f_n - g_n|| < \delta\epsilon$. \\ \ \\

Since $f_n$ is known to be of order $\epsilon$, thus when its difference with $g_n$ is arbitrarily smaller than $\epsilon$, one stays as a good approximation of the other. \\ \ \\

{\bf Proof:} First of all, we define
\begin{equation}
\Delta_n \equiv c\epsilon^l \sum_{i=0}^{n-1}(1+b\epsilon^{m-1})^i
=c\epsilon^l \frac{(1+b\epsilon^{m-1})^n-1}{b\epsilon^{m-1}}~.
\label{eq-RecClose}
\end{equation}
Within the range of $n$ stated in the Theorem, it is easy to see that
\begin{equation}
\Delta_n \leq \Delta_{\lfloor \alpha\epsilon^{1-m} \rfloor}<
c\epsilon^l 
\frac{(1+b\epsilon^{m-1})^{\alpha\epsilon^{1-m}}-1}
{b\epsilon^{m-1}}
<\frac{c\epsilon^{1+l-m}}{b}
\left(e^{b\alpha}-1\right)~.
\end{equation}
Since $l>m$, there is always a choice of $\epsilon$ such that $\Delta_n<\delta\epsilon$. We will choose an $\epsilon$ small enough for that, and also small enough such that
\begin{eqnarray}
||f_n|| + a\epsilon^m &<& \epsilon + a\epsilon^m < r~, \\     
||f_n|| + \Delta_n &<& \epsilon + \delta\epsilon < r~.
\end{eqnarray}

Next, we use mathematical induction to prove that given such choice of $\epsilon$,
\begin{equation}
||f_n - g_n|| \leq \Delta_n~.
\end{equation}

For $n=0$, this is trivially true.
\begin{equation}
||f_0 - g_0|| = 0 \leq \Delta_0 = 0~.
\end{equation}

Assume this is true for $(n-1)$,
\begin{equation}
||f_{n-1} - g_{n-1}|| \leq \Delta_{n-1}~.
\end{equation}
Combine it with Eq.~(\ref{eq-FGclose}) and (\ref{eq-Lsmooth}), we can derive for the next term in the sequence.
\begin{eqnarray}
||f_n - g_n|| &\leq& 
||f(f_{n-1}) - g(f_{n-1})|| + ||g(f_{n-1})- g(g_{n-1})||
\\ \nonumber
&\leq& c\epsilon^l + (1+b\epsilon^{m-1})\Delta_{n-1}
= \Delta_n~.
\end{eqnarray}
Thus by mathematical induction, we have proven the theorem. 

Note that although in the early example for physical intuitions, we took $f$ as the full nonlinear theory and $g$ as the truncated theory, their roles are actually interchangeable in Theorem I. Thus we can use the theorem in both ways. If a fully nonlinear solution, presumably obtained by numerical methods, stays below $\epsilon$, then Theorem I guarantees that a truncated theory can reproduce such solution. The reverse is also true. If the truncated theory leads to a solution that stays below $\epsilon$, then Theorem I guarantees that this is a true solution reproducible by numerical evolution of the full nonlinear theory.

Also note that the truncated theory might belong to an expansion which does not really converge to the full nonlinear theory. This is quite common in field theories that a na\"ive expansion is only asymptotic instead of convergent. Theorem I does not care about whether such full expansion is convergent or not. It only requires that the truncated theory is a good approximation to the full theory up to some specified order, as stated in Eq.~(\ref{eq-FGclose}). Divergence of an expansion scheme at higher orders does not invalidate our result.\footnote{We thank Jorge Santos for pointing out the importance to stress this point.}

Finally, if one takes a closer look at Eq.~(\ref{eq-RecClose}), one can see that if $n$ is allowed to be larger than the $\epsilon^{1-m}$ time scale, for example $n\sim\epsilon^{-s}$ with $s>m-1$, then $\Delta_n$ fails to be bounded from above in the $\epsilon\rightarrow0$ limit. Since the upperbound we put is already quite optimal, we believe that the truncated theory does break down at any longer time scale. In particular, this does not care about $l$. Namely, independent of how small the truncated error is, accumulation beyond the $\epsilon^{1-m}$ time scale always makes the truncated dynamic a bad approximation for the full theory. Thus, the conventional wisdom only requires a small correction. {\it Usually, the truncated theory breaks down at the $\epsilon^{1-m}$ time scale. Occasionally, it can still hold at exactly this time scale but breaks down at any longer time scale.}


\section{Theorem II: Conditionally Preserved Symmetry}
\label{sec-thm2}

We will consider an example that the truncated theory has an approximate scaling symmetry. Let $L(x)=x$, $g(x) = L(x) + G(x)$, such that for all $||x||, ||y||<r$,
\begin{eqnarray}
||G(x)|| &<& a||x||^m~, \\
||G(x)-G(y)|| &<& b\cdot||x-y||{\rm Max}(||x||,||y||)^{m-1}~, \\
||G(x) - N^m G(x/N)|| &<& d||x||^p~,
\end{eqnarray}
for a given $p>m$ and any $N>1$. Namely, the linear theory is trivial that $L_n=x$ does not evolve with $n$. The only evolution for $g_n$ comes from the function $G(x)$, which is for many purposes effectively ``an $x^m$ term''. In this case, it is reasonable to expect a rescaling symmetry: reducing the amplitude by a factor of $N$, but evolve for a time longer by a factor of $N^{m-1}$, leads to roughly the same result. \\ \ \\

{\bf Theorem II:} For any finite $\delta>0$ and $\alpha>0$, there exists $0<\epsilon<r$ such that if $||g_n(x)||<\epsilon~$ for all $0\leq n <\alpha\epsilon^{1-m}$, then 
\begin{eqnarray}
||Ng_n(x/N)- (1-\beta) g_{n'}(x) - \beta g_{n'+1}(x)|| < \delta\epsilon~.
\end{eqnarray}  
Here $n' = \lfloor (nN^{1-m}) \rfloor$ is the largest integer smaller than or equal to $(nN^{1-m})$, and $\beta = (nN^{1-m})-n'$. This should be valid for any $N>1$ and for $0\leq n < \alpha(\epsilon/N)^{1-m}$. \\ \ \\

The physical intuition is the following. Every term in the rescaled sequence stays arbitrarily close to some weighted average of the terms in the original sequence, which exactly corresponds to the appropriate ``time'' (number of steps) of the rescaling. We will first prove this for a special case, $N=2^{\frac{1}{m-1}}$. This case is particularly simple, since such rescaling exactly doubles the length of the sequence, and $\beta$ will be either $0$ or $1/2$ which leads to two specific inequalities to prove:

\begin{eqnarray}
\bigg|\bigg| 2^{\frac{1}{m-1}}g_{2n-1}(x/2^{\frac{1}{m-1}}) - 
\frac{g_{n-1}(x) + g_n(x)}{2}\bigg|\bigg| 
&\leq& C \cdot\epsilon^q~, \label{eq-half1} \\
\bigg|\bigg| 2^{\frac{1}{m-1}}g_{2n}(x/2^{\frac{1}{m-1}}) - 
g_n(x)\bigg|\bigg| 
&\leq& C \cdot\epsilon^q~,
\label{eq-half2}
\end{eqnarray}
for some $C>0$ and $q>1$. This will again be done by a mathematical induction.

During the process, it should become clear that the proof can be generalized to any $N>1$. We will not present such proof, because the larger variety of $\beta$ values makes it more tedious, although it is still straightforward. However, for the self-completeness of this paper, what we need in the next session is only that $N$ can be arbitrarily large. Through another mathematical induction, we can easily prove Theorem II for $N=2^{\frac{k}{m-1}}$ for arbitrarily positive integer $k$. It is still a bit tedious, so we will present that in Appendix \ref{sec-Nk}. \\ \ \\

{\bf Proof for $ \mathbf{ N=2^{\frac{1}{m-1}}}$ }  \\

We start by defining the monotonically increasing function

\begin{eqnarray}
\Delta_n = \left( \frac{d}{b}2^{m-1} \epsilon ^{p-m+1} + \frac{a}{2}  \epsilon ^m \right) \left[ \left( 1 + \frac{b}{2} (2 \epsilon) ^{m-1} \right)^{n} -1 \right],
\end{eqnarray}
with the properties
\begin{eqnarray}
\Delta _n & < & \Delta _{\alpha (\epsilon/N)^{1-m}} 
\label{eq:bound} \\
& = & \left( \frac{d}{b}2^{m-1} \epsilon ^{p-m+1} + \frac{a}{2}  \epsilon ^m \right) \left[ \left( 1 + \frac{b}{2} (2 \epsilon) ^{m-1} \right)^{\alpha (\epsilon / N)^{1-m}} -1 \right]  \nonumber \\
          & < & \left( \frac{d}{b}2^{m-1} \epsilon ^{p-m+1} + \frac{a}{2} \epsilon ^m \right) \left[e^{ 2^{m-1} \alpha b/4} -1 \right]  \label{eq:q}  <  C \cdot \epsilon^{q}~,  \nonumber \\
\Delta _{n+1} & = & \Delta _n \left(1 + \frac{b}{2}(2 \epsilon)^{m-1}\right) + \frac{d}{2} \epsilon ^p + \frac{ab}{4} \epsilon ^m (2\epsilon) ^{m - 1}~.
\label{eq-RecRescale}
\end{eqnarray}
The meaning of Eq.~(\ref{eq:bound}) is that, in the range we care about, $\Delta _n$ is bounded from above by a power of $\epsilon$ higher than one, since $q =$Min$\{p-m+1,m\}$. Given that, we can always choose $\epsilon$ small enough such that
\begin{eqnarray}
\big| \big| g_n (x) \big| \big| + \Delta _n & < & \epsilon + C \cdot \epsilon ^q < r~.
\end{eqnarray}

Given our choice of $\epsilon$, we can employ mathematical induction to prove that
\begin{eqnarray}
\Big| \Big| N g_{2n - 1} (x/N) - \frac{g_{n-1}(x) + g_{n}(x)}{2} \Big| \Big| & \leq & \Delta _{2n-1} \\ 
\Big| \Big| Ng_{2n}(x/N) - g_n (x) \Big| \Big| & \leq & \Delta _{2n}~, \label{eq-even}
\end{eqnarray}
which prove Eq.~(\ref{eq-half1}) and (\ref{eq-half2}).

First, we observe that for $n=0$,
\begin{eqnarray}
\Big| \Big|  Ng_{0}(x/N) - g_0 (x) \Big| \Big|  = \Big| \Big|  n \frac{x}{N} - x \Big| \Big|  = 0 < C \cdot \epsilon ^q~,
\end{eqnarray} 
is obviously true. Then, we assume that Eq.~(\ref{eq-even}) is true for $n$ in the original sequence and $2n$ in the rescaled sequence. We can prove for the next term, the $(2n+1)$ term in the rescaled sequence. 
\begin{eqnarray}
& \vphantom{=}& \Big| \Big| Ng_{2n+1}(x/N) - \frac{g_{n}(x) + g_{n+1}(x)}{2} \Big| \Big| \nonumber \\
                                                          & = & \Big| \Big| Ng_{2n}(x/N) + NG \big( g_{2n}(x/N) \big) - g_n(x) - \frac{1}{2}G\big( g_n(x) \big) \Big| \Big| \nonumber \\
                                                          & = & \Big| \Big| Ng_{2n}(x/N) + NG \big( g_{2n}(x/N) \big) - g_n(x) - \frac{1}{2}G\big( g_n(x) \big) \nonumber \\ 
                                                          & + & NG\big(g_n(x)/N\big)  - NG\big(g_n(x)/N\big) \Big| \Big|   \nonumber \\
                                                          & < &\Big| \Big| Ng_{2n}(x/N) - g_n(x) \Big| \Big| + N \Big| \Big| G \big( g_{2n}(x/n) \big) -  G\big(g_n(x)/N\big) \Big| \Big| \nonumber \\ 
                                                          & + & \frac{1}{N^{m-1}} \Big| \Big| G\big(g_n(x)\big) - N^m G \big( g_n(x)/N \big) \Big| \Big| \nonumber \\
                                                          & < & \Delta _{2n} + \Delta_{2n}\frac{b}{N^{m-1}} \left( \big| \big| \Delta _{2n} \big| \big| + \big| \big| g_n(x) \big| \big| \right)^{m-1} + \frac{d}{2} \epsilon ^p \nonumber \\
                                                          & < & \Delta _{2n}  + \Delta _{2n}\frac{b}{2} \left( \Delta _{2n}  +  \epsilon \right)^{m-1} + \frac{d}{2} \epsilon ^{p} \nonumber \\
                                                          & < & \Delta _{2n}  + \Delta _{2n}\frac{b}{2} (  2 \epsilon) ^{m-1} + \frac{d}{2} \epsilon ^{p}  <  \Delta _{2n +1}.
\end{eqnarray}
Similarly we can prove for the $(2n+2)$ term in the rescaled sequence, which is the $(n+1)$ term in the original sequence.
\begin{eqnarray}
& \vphantom{=} & \Big| \Big| N g_{2n + 2}(x/N) - g_{n + 1}(x) \Big| \Big| \nonumber \\
                                             & = & \Big| \Big| N g_{2n + 1}(x/N) + NG\big( g_{2n+1}(x/N) \big) - \frac{1}{2} g_{n + 1}(x) \nonumber \\ 
                                             & - & \frac{1}{2} g_{n}(x) - \frac{1}{2} G\big(g_n(x)\big) + NG \left( \frac{g_n(x) + g_{n+1}(x)}{N^m} \right) - NG \left( \frac{g_n(x) + g_{n+1}(x)}{N^m} \right) \nonumber \\
                                             & + & NG \left( \frac{g_n(x)}{N} \right) - NG \left( \frac{g_n(x)}{N} \right) \Big| \Big| \nonumber \\
                                             & < & \Big| \Big|N g_{2n+1}(x/N) - \frac{g_{n+1}(x) + g_n(x)}{2} \Big| \Big| 
                                              +  N \Big| \Big| G\big( g_{2n+1}(x/N) \big) -  G \left( \frac{g_n(x) + g_{n+1}(x)}{N^m} \right)  \Big| \Big| \nonumber \\
                                             & + & N \Big| \Big| G \left( \frac{g_n(x) + g_{n+1}(x)}{N^m} \right) - G \left( \frac{g_n(x)}{N} \right)  \Big| \Big|  +  N \Big| \Big| G \left( \frac{g_n(x)}{N} \right) - \frac{1}{N^{m-1}} G\big(g_n(x)\big) \Big| \Big| \nonumber \\
                                             & < & \Delta _{2n + 1} + \Delta _{2n+1} \frac{b}{2} ( 2 \epsilon ) ^{m-1} + \frac{d}{2} \epsilon ^p  + \frac{ab}{4} \epsilon ^m ( 2 \epsilon)  ^{m - 1} = \Delta _{2n+2}~.
\end{eqnarray}
This completes the mathematical induction.

Eq.~(\ref{eq-RecRescale}) takes basically the same form as Eq.~(\ref{eq-RecClose}). Thus, Theorem II also only holds up to exactly the $\epsilon^{1-m}$ time scale, but not any longer.


\section{Application: Persistence of Stability Islands}
\label{sec-island}

First, we review the ``stability island conjecture'' argued by Dias, Horowitz, Marolf and Santos in \cite{DiaHor12}. Numerical simulations suggest that given a small but finite initial amplitude $\phi_{\rm init.}\sim\epsilon$ in AdS space with Dirichlet boundary condition, dynamical evolution can lead to black hole formation at the long time scale $T\sim\epsilon^{-2}$ \cite{BizRos11}\footnote{Note that for this purpose, $m=3$, thus $\epsilon^{-2}$ is the relevant time scale.}. In the meanwhile, some initial conditions do not lead to black holes at the same time scale. In particular, there are special solutions (set of measure zero) which not only do not collapse, they stay exactly as they are. These especially stable solutions are called geons (in pure gravity) or Boson-stars/Oscillons (scalar field)\cite{DiaHor11,BucLie13,MalRos13a}\footnote{There are also quasi-periodic solutions which do not stay exactly the same but demonstrate a long-term periodic behavior and the energy density never gets large \cite{BalBuc14}.}. At finite amplitudes, they are not only stable themselves, but also stabilize an open neighborhood in the phase space, forming stability islands which prevent nearby initial conditions from collapsing into black holes in the $\sim\epsilon^{-2}$ time scale.

\begin{figure}[tb]
\begin{center}
\includegraphics[width=6cm]{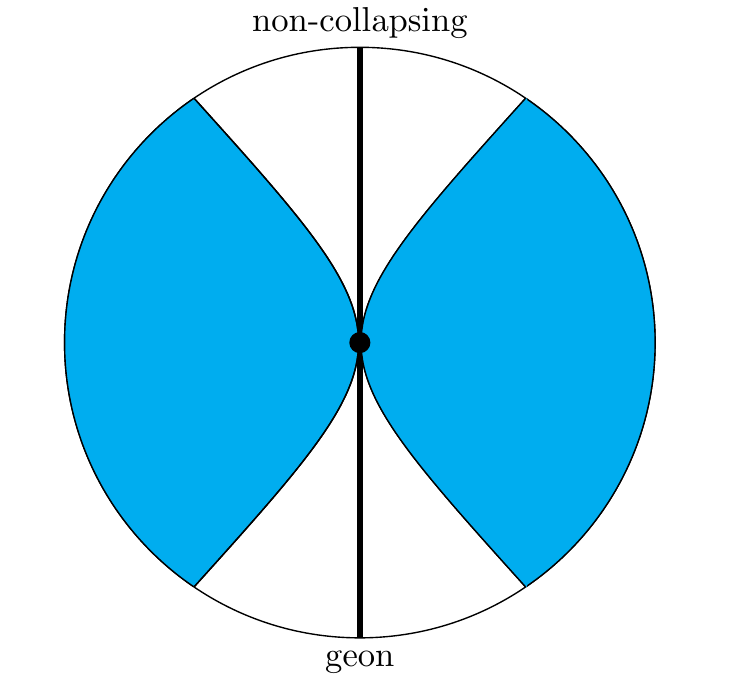}
\includegraphics[width=6cm]{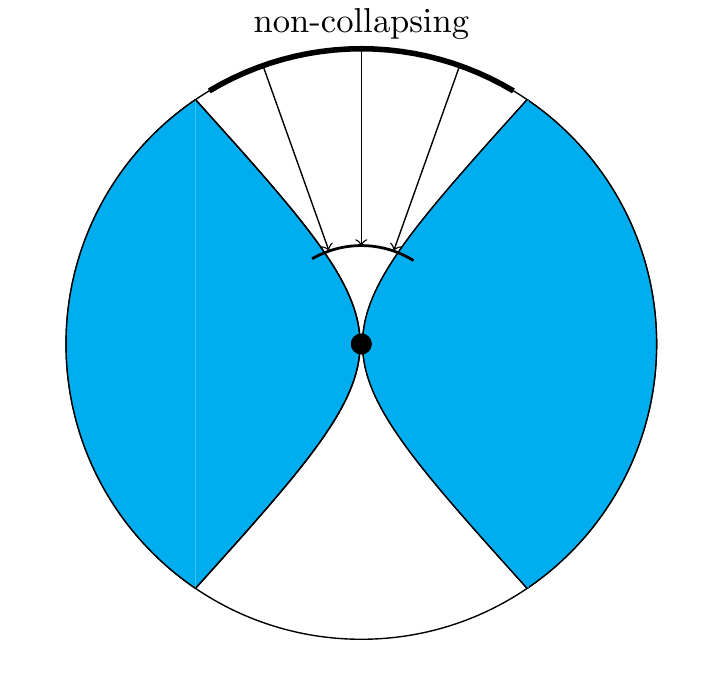}
\caption{Phase space diagrams of small initial perturbations around empty AdS (central black dot) according to the stability island conjecture. The radial direction represents field amplitude (total energy), and the angular direction represents field profile shape (energy distribution). Initial perturbations in the shaded (blue) region will collapse into black holes at the $\sim\epsilon^{-2}$ time scale, while the unshaded region, around the exactly stable geons (thick black line), will not. The unshaded region is cuspy, showing that according to the conjecture, the angular span of non-collapsing perturbations goes to zero as amplitude goes to zero. The right panel demonstrate the usage of both Theorems we proved in this paper. We can transport the known, non-collapsing solutions, directly in the radial direction, to an arc of identical angular span at an arbitrarily smaller radius. It is a direct contradiction to the cuspy nature of the unshaded region.
\label{fig-island}}
\end{center}
\end{figure}

Dias, Horowitz, Marolf and Santos argued that such stabilization effect can be understood as breaking the AdS resonance.\footnote{In some sense, this argument \cite{DiaHor12} provides a stronger support for non-collapsing solutions to have a nonzero measure, because it goes beyond spherical symmetry. Current numerical results are limited to spherical symmetry, thus strictly speaking cannot establish a nonzero measure for either collapsing or non-collapsing results. This is why controversies over some numerical results \cite{BizRos14,BucGre15} should not undermine the believe that non-collapsing solutions form a set of nonzero measure at finite amplitude.} It should lose strength as the geon's own amplitude decreases. Thus, such stability islands go away in the limit of zero amplitude. The easiest way to summarize their conjecture is by the cuspy phase-space diagram in Fig.\ref{fig-island}. Other than the measure-zero set of exact geons/Boson-stars, non-collapsing solutions at finite amplitude will all end up collapsing as amplitude goes to zero.

Next, we will show that the requirements of both Theorem I and II are applicable to the AdS (in)stability problem. For simplicity, we present the analysis on a massless scalar field in global AdS space of Dirichlet boundary condition. Metric fluctuation in pure gravity will also meet the requirements\cite{DiaHor12,HorSan14}. We will avoid going into specific details of the AdS dynamics, but only provide the relevant papers where those can be found.
\begin{itemize}
\item The linear space $\mathcal{H}$ we used to state both Theorems (see the beginning of Sec.\ref{sec-thm1}) contains all smooth functions $\phi(\vec{r})$ on the domain of the entire spatial slice of the AdS space at one global time between.
\item The function $L$ evolves one such function forward for one ``$AdS$ period'', namely $T = 2\pi R_{\rm AdS}$ in the explanation right below Eq.~(\ref{eq-FGclose}), using the fixed background equation of motion. It includes no gravitational self-interaction and is a linear function. Actually, since the AdS spectrum has integer eigenvalues, the evolution is exactly periodic \cite{HamKab06,HamKab06a}. $L(x)=x$ is trivial, automatically conserves length and also meets the requirement for Theorem II.
\item The definition of the norm is trickier. We first evolve $x$, using the fixed-background evolution, for exactly $2\pi R_{\rm AdS}$, and examine the maximum local energy density ever occurred during such evolution. The norm is defined to be the square-root of this value, $\sqrt{\rho_{\rm Max}}$. The evolution is linear, and the quantity is both a maximum and effectively an absolute value, thus it satisfies the triangular inequality.\footnote{The reason why we adopt this tortuous definition of norm is to guarantee that gravitational interaction during one AdS time stays weak when the norm is small, thus we can apply both theorems. Note that defining total energy as the norm would fail such purpose.}
\item The actual dynamic, including Einstein equations, is clearly nonlinear. When the maximum energy density is always small, the gravitational back-reaction is well-bounded. One can perform a recursive expansion in which the leading order correction to the linear dynamic comes from coupling to its own energy, $\rho\phi \propto \phi^3$ \cite{BizChm08,BizRos11,DFLY14}. A theory truncated at this order and the full nonlinear theory can be our $f$ and $g$, interchangeably, in Theorem I with $m=3$. \footnote{Such expansion, continued to higher orders, is likely only asymptotic instead of convergent. As explained in Sec.\ref{sec-thm1}, that does not cause a problem for our theorems.}
\item The $\phi^3$ contribution calculated in different approximation methods might be different\cite{BizChm08,BizRos11,DFLY14}, but they all satisfy the approximate rescaling symmetry required for the function $G$ in Theorem II.
\end{itemize}

Now we have established the applicability of both Theorems in this paper, the stability island conjecture can be disproved in three simple steps.
\begin{enumerate}
\item At a small but finite amplitude where measure nonzero sets of non-collapsing solutions exist (the outermost thick arc in Fig.\ref{fig-island}), apply Theorem I to translate them into solutions in the truncated theory.
\item Use Theorem II to scale down the above solutions to arbitrarily small amplitudes. That means projecting radially in Fig.\ref{fig-island} into an arc of the same angular span.
\item Use Theorem I again to translate these rescaled solution in the truncated theory back to the full nonlinear theory. This establishes the existence of non-collapsing solutions as a set of measure non-zero (an arc of finite angular span in Fig.\ref{fig-island}). \footnote{The only works for rescaling down to smaller amplitudes. Rescaling to larger amplitudes can easily exceed the radius of validity of perturbative expansion even at short time scales.}
\end{enumerate}
Thus, we have established that the measure-non-zero neighborhood stabilized by a geon at finite amplitude, if never evolves to high local energy density during the long time scale, directly guarantees the same measure-non-zero, non-collapsing neighborhood at arbitrarily smaller amplitudes. This directly contradicts the stability island conjecture.

It is interesting to note that the collapsing solutions always have energy density large at a certain point, thus neither theorems we proved here are applicable to them. As a result, one cannot establish their existence at arbitrarily small amplitudes through a similar process. Therefore, the opposite possibility to the stability island conjecture, that collapsing solutions disappear into a set of measure zero at zero amplitude, is still consistent with current evidence.

\acknowledgments

We thank the hospitality of Kavli Institute for Theoretical Physics at Santa Barbara during the workshop of ``Quantum Gravity Foundation, From UV to IR''. In particular, we are inspired by the discussion with Ben Freivogel, Don Marolf and Gary Horowitz during our time in the workshop. We are also grateful for the discussions with Piotr Bizon, Alex Buchel, Matt Lippert, Luis Lehner and Andrzej Rostworowski.

\appendix

\section{Arbitrarily small rescaling}
\label{sec-Nk}

In Sec.\ref{sec-thm2}, we have proven Theorem II for $N=2^{\frac{1}{m-1}}$. Now, we will generalize it to arbitrary $N'=2^{\frac{k}{m-1}} = N^k$, for any $ k \in {\mathbb N}^{+} $ : 
\begin{eqnarray}
\Big| \Big| N^k g_{n}(x/N^k) - \left( 1 - \beta _{k}(n) \right) g_{ \floor*{\frac{n}{2^k}}}(x) - \beta _k(n)  g_{ \floor*{\frac{n}{2^k}} + 1 }(x) \Big| \Big| \leq C' \cdot \epsilon ^q, \label{eq:theorem}
\end{eqnarray}
where we have written down explicitly the dependence of $\beta$ on $k$ and $n$:
\begin{eqnarray}
\beta_{k}(n) = \frac{2}{n^k} - \floor*{\frac{n}{2^k}} ,  
\end{eqnarray}
which possesses the following properties for positive integers $j$ and $l$:
\begin{eqnarray}
\beta _{k+1}(2l)   & = & \beta _{k}(l)~, \ \ \ \ \ \ \ \ \ \ \ \ \ 
\ \ \ \ \ \ \ 
{\rm this \ is \ always \ true}~;\\
\beta _{k+1}(2l+1) & = & \frac{1}{2} \beta _k (l) + \frac{1}{2} \beta _k (l+1) \label{eq:beta}, \quad \text{for } l + 1 \neq j \cdot 2^k ~;  \\
\beta _{k+1}(2l+1) & = & \frac{1}{2} \beta _k (l) + \frac{1}{2} [ 1 - \beta _k (l+1) ] \label{eq:betaa}, \quad \text{for } l + 1 = j \cdot 2^k~. 
\end{eqnarray}

These follow naturally from the properties of the floor function that
\begin{eqnarray}
\floor*{\frac{2l + 1}{2^{k+1}}} & = & \floor*{\frac{l}{2^k}}~, 
\ \ \ \ \ \ \ \ \ {\rm is \ always \ true}~,
\end{eqnarray}
and
\begin{eqnarray}
\floor*{\frac{2l + 1}{2^{k+1}}} & = & \floor*{\frac{l + 1 }{2^k}}, \quad \text{when} \quad l \neq j \cdot 2^k - 1~; \\
\floor*{\frac{2l + 1}{2^{k+1}}} & = & \floor*{\frac{l + 1 }{2^k}} - 1 , \quad \text{when} \quad l = j \cdot 2^k - 1~.
\end{eqnarray}
%
%
%
%
%
%
We now define:
\begin{eqnarray}
F_k \equiv C \cdot \epsilon ^q \sum _{i = 0}^{k} N^{i(1-q)} = \frac{C \cdot \epsilon ^q}{1 - N^{1-q}} \left( 1 - N^{k(1-q)} \right) \leq C' \cdot \epsilon ^q~,
\end{eqnarray}
for $C' = C/(1-N^{1-q})$. This converges as $k \rightarrow \infty$, since $1-q < 0 $, and satisfies the recursive relation: 
\begin{eqnarray}
F_{k+1} & = & F_k + C \cdot N^{k(1-q)}.
\end{eqnarray}
Now we will prove Eq.~(\ref{eq:theorem}) by induction. We have already shown that it holds for $k=1$ in Sec.\ref{sec-thm2}, hence assuming that it holds for arbitrary $k$, we want to show that it holds for $k+1$ as well. \\

It is helpful to split the proof in three parts, one for $n=2l$, one for $n=2l + 1$, with $l \neq j \cdot 2^k - 1$ and one for $n = 2l + 1$, with $l= j \cdot 2^k - 1$. 
\begin{enumerate}

\item \textit{Part 1: $n= 2l$}

\begin{eqnarray}
& \vphantom{=} & \Big| \Big| N^{k+1} g_{2l}(x/N^{k+1}) - \left( 1 - \beta _{k+1}(2l) \right) g_{ \floor*{\frac{2l}{2^{k+1}}}}(x) - \beta _{k+1}(2l)  g_{ \floor*{\frac{2l}{2^{k+1}}} + 1 }(x) \Big| \Big| \nonumber \\
                    & = &  \Big| \Big| N \cdot N^{k} g_{2l}\left(\frac{x/N^{k}}{N} \right) - \left( 1 - \beta _{k}(l) \right) g_{ \floor*{\frac{2l}{2^{k+1}}}}(x) - \beta _{k}(l)  g_{ \floor*{\frac{2l}{2^{k+1}}} + 1 }(x) \Big| \Big| \nonumber \\
                     & < & N^k \Big| \Big| N g_{2l} \left( \frac{x/N^k}{N} \right) - g_l \left( x/N^k \right) \Big| \Big| \nonumber \\ 
                     & + &  \Big| \Big| N^k g_l \left( x/N^k \right) - \big( 1 - \beta _k (l) \big) g_{ \floor*{\frac{l}{2^{k}}}}(x)  - \beta _k (l)g_{ \floor*{\frac{l}{2^{k}}} + 1 }(x)  \Big| \Big| \nonumber \\
                     & < & C \cdot N^{k(1-q)} + F_k = F_{k+1}. 
\end{eqnarray}

\item \textit{Part 2: $n= 2l + 1$, with $l \neq j \cdot 2^k - 1$}

\begin{eqnarray}
& \vphantom{=} & \Big| \Big| N^{k+1} g_{2l + 1}(x/N^{k+1}) - \left( 1 - \beta _{k+1}(2l + 1) \right) g_{ \floor*{\frac{2l + 1}{2^{k+1}}}}(x) - \beta _{k+1}(2l + 1)  g_{ \floor*{\frac{2l + 1}{2^{k+1}}} + 1 }(x) \Big| \Big| \nonumber \\ 
                      & = & \Big| \Big| N \cdot N^{k} g_{2l + 1}\left(\frac{x/N^{k}}{N} \right) - \left( 1 - \frac{1}{2}\beta _{k}(l) - \frac{1}{2} \beta _k (l + 1) \right) g_{ \floor*{\frac{2l + 1}{2^{k+1}}}}(x) \nonumber \\ 
                      & - & \frac{1}{2} \left( \beta _{k}(l) + \beta _k(l+1) \right) g_{ \floor*{\frac{2l + 1}{2^{k+1}}} + 1}(x) \Big| \Big|  \nonumber \\
                      & < & N^k \Big| \Big| N g_{2l + 1} \left( \frac{x/N^k}{n} \right) - \frac{g_l(x/N^k) + g_{l+1}(x/N^k)}{2} \Big| \Big| \nonumber \\ 
                      & + & \frac{1}{2} \Big| \Big| N^k g_l (x/N^k) - \left( 1 - \beta _k (l) \right)g_{ \floor*{\frac{2l + 1}{2^{k+1}}}}(x) - \beta _k (l) g_{ \floor*{\frac{2l + 1}{2^{k+1}}} + 1}(x)  \Big| \Big|  \nonumber \\ 
                      & + &  \frac{1}{2} \Big| \Big| N^k g_{l+1} (x/N^k) - \left( 1 - \beta _k (l + 1) \right)g_{ \floor*{\frac{2l + 1}{2^{k+1}}}}(x) - \beta _k (l+1) g_{ \floor*{\frac{2l + 1}{2^{k+1}}} + 1}(x)  \Big| \Big| \nonumber \\ 
                      & < &  C \cdot N^{k(1-q)} + 2 \frac{1}{2} F_k  
											= F_{k+1}. 
\end{eqnarray}

\item \textit{Part 3: $n= 2l + 1$, with $l = j \cdot 2^k - 1$}

\begin{eqnarray}
& \vphantom{=} & \Big| \Big| N^{k+1} g_{2l + 1}(x/N^{k+1}) - \left( 1 - \beta _{k+1}(2l + 1) \right) g_{ \floor*{\frac{2l + 1}{2^{k+1}}}}(x) - \beta _{k+1}(2l + 1)  g_{ \floor*{\frac{2l + 1}{2^{k+1}}} + 1 }(x) \Big| \Big|  \nonumber \\ 
                          & = & \Big| \Big| N \cdot N^{k} g_{2l + 1}\left(\frac{x/N^{k}}{N} \right) - \left( 1 - \frac{1}{2}\beta _{k}(l) - \frac{1}{2} \big( 1 - \beta _k (l + 1) \big) \right) g_{ \floor*{\frac{2l + 1}{2^{k+1}}}}(x) \nonumber \\
                          & - &  \frac{1}{2} \left( \beta _{k}(l) + \big(1 - \beta _k(l+1) \big) \right) g_{ \floor*{\frac{2l + 1}{2^{k+1}}} + 1}(x) \Big| \Big| \nonumber \\ 
                          & < & N^k \Big| \Big| N g_{2l + 1} \left( x/N^k \right) - \frac{g_l (x/N^k) + g_{l+1}(x/N^k)}{2}  \Big| \Big| \nonumber \\
                          & + & \frac{1}{2} \Big| \Big| N^k g_l (x/N^k) - \beta _k (l) g_{ \floor*{\frac{l}{2^{k}}} + 1 }(x) - \big( 1 - \beta _k (l)\big)g_{ \floor*{\frac{l}{2^{k}}}}(x)  \Big| \Big|  \nonumber \\ 
                          & + & \frac{1}{2} \Big|  \Big| N^k g_{l+1} (x/N^k) - \beta _k(l+1) g_{ \floor*{\frac{l + 1}{2^{k}}} - 1 }(x) - \big( 1 - \beta _k (l+1) \big)g_{ \floor*{\frac{l + 1}{2^{k}}} }(x) \Big| \Big| \nonumber \\ 
                          & < &  C \cdot N^{k(1-q)} + 2 \frac{1}{2} F_k = F_{k+1}.
\end{eqnarray}
We have used here the fact:
\begin{eqnarray}
\beta _k(l+1) g_{ \floor*{\frac{l + 1}{2^{k}}} - 1 }(x) = \beta _k(l+1) g_{ \floor*{\frac{l + 1}{2^{k}}} + 1 }(x),
\end{eqnarray}
since $\beta _k (j \cdot 2^k) = 0$. 

\end{enumerate}

\bibliographystyle{utcaps}
\bibliography{all_active}

\end{document}